\documentclass[aps,pra,twocolumn,showpacs,floatfix]{revtex4-1}
\usepackage{graphicx}

\begin{document}
\title{Optically induced conical intersections in traps
for ultracold atoms and molecules}
\author{Alisdair O. G. Wallis}
\affiliation{Department of Chemistry, Durham University, South
Road, Durham, DH1~3LE, United Kingdom}
\author{Jeremy M. Hutson}
\affiliation{Department of Chemistry, Durham University, South
Road, Durham, DH1~3LE, United Kingdom}

\date{\today}

\begin{abstract}
We show that conical intersections can be created in laboratory
coordinates by dressing a parabolic trap for ultracold atoms or
molecules with a combination of optical and static magnetic
fields. The resulting ring trap can support single-particle
states with half-integer rotational quantization and
many-particle states with persistent flow. Two well-separated
atomic or molecular states are brought into near-resonance by
an optical field and tuned across each other with an
inhomogeneous magnetic field. Conical intersections occur at
the nodes in the optical field.
\end{abstract}


\maketitle

There is great interest in the properties of ultracold atoms in traps
with unusual shapes, especially in rings and other multiply connected
geometries. Both magnetic \cite{Gupta:2005, Arnold:2006,
Heathcote:2008, Baker:2009} and optical \cite{Olson:2007} ring traps
have been created. Such traps have potential uses in inertial sensing
and in atom interferometry. They also exhibit geometric phases (Berry
phases) due to the spatial variation in the magnetic field direction
\cite{Ho:vortex:1996, Olshanii:1998, Zhang:ring:2006}. These quantum
phases may be exploited to create neutral-atom analogues of systems
involving charged particles in magnetic fields \cite{Olshanii:1998}.
Lin {\em et al.} \cite{Lin:2009} have recently used Berry phases
generated with a spatially varying Raman coupling to create large
synthetic magnetic fields, with potential for studying phenomena such
as the quantum Hall effect and topological quantum computing.

Berry phases of a qualitatively different type would be be accessible
in a ring trap with a conical intersection (CI) at the center. A CI
occurs when two adiabatic potential energy surfaces intersect, at a
point in two dimensions or on a surface of dimension $(n-1)$ in $n$
dimensions. The surfaces have the topology of a double cone near the
intersection. A CI produces a {\em half-integer} Berry phase
\cite{herzberg:1963, Berry:1984}: the internal wavefunction of a state
that fully encircles the intersection adiabatically changes sign upon
completion of a circuit around it.

CIs can exist in the internal vibrational coordinates of polyatomic
molecules, \cite{domcke:conical:2004}, where they have important
consequences for collision dynamics and photochemistry, and in momentum
space, where they have important consequences for electronic band
structure and conductivity. CIs in momentum space are often referred to
as \emph{Dirac cones}, and can occur both in solid-state systems such
as graphene \cite{Novoselov:graphene:2005, Zhang:graphene:2005} and in
cold atoms in specific laser and optical lattice configurations
\cite{Dudarev:2004, Zhu:2007, Larson:2009}. Moiseyev {\em et al.}\
\cite{Moiseyev:laserCI:2008} have shown that arrays of CIs may occur in
a combination of molecular rovibrational coordinates and spatial
coordinates for diatomic molecules in optical lattices, and may have
important consequences for the dynamics of ultracold molecules in
standing laser waves. W\"uster {\em et al.}\ \cite{Wuster:2011} have
proposed creating CIs with groups of ultracold atoms or molecules
interacting through resonant dipole interactions.

We have recently shown that, for a gas of ultracold molecules with both
electric and magnetic dipoles, CIs can be created purely in laboratory
position space with externally applied static electric and magnetic
fields \cite{Wallis:PRL:CI:2009}. The magnetic field is used to produce
a crossing between two states of opposite parity, and the electric
field breaks the parity symmetry and induces an avoided crossing
between the states \cite{Friedrich:2000, Abrahamsson:2007}. An CI is
created where the electric field passes through zero. If the
intersection is surrounded by an optical dipole trap, the resulting
adiabatic potential has a toroidal minimum that encircles the CI. The
geometric phase effect can then produce a ground state with persistent
flow and half-integer quantization of the angular momentum for rotation
around the ring.

CIs of the type described in ref.\ \cite{Wallis:PRL:CI:2009} are
unlikely to be accessible for atoms, because there are no coolable
atomic systems that have two states of opposite parity close enough
together to be brought into degeneracy by a magnetic field. However, a
stable degenerate gas of molecules with both electric and magnetic
dipole moments is also some way away. The object of the present paper
is to propose an alternative that is feasible for atoms, in which two
states that cross as a function of magnetic field are coupled by a
microwave or laser field. For generality, we refer to the microwave or
laser field as an optical field. The coupling is proportional to the
amplitude of the optical field, so CIs occur at nodes in the field,
where the amplitude is zero. This approach has the advantage that the
two states to be coupled do not need to be near-degenerate before the
optical field is applied. In addition, it allows CIs to be created
between pairs of states of the {\em same} parity, which are coupled by
the magnetic component of the optical field.

CIs of the type proposed here will make it possible to study new types
of interference effect in ultracold atomic systems. In addition to the
half-integer quantization and persistent flow discussed in ref.\
\cite{Wallis:PRL:CI:2009}, it will be possible to explore dynamical
effects produced by the Berry phase. For example, if portions of a
matter wave pass either side of the CI, the two parts will interfere
destructively when they meet on the far side. If two or more CIs can be
produced in the same trap, as described below, even richer
interferences will be possible between paths that encircle one CI and
change sign and paths that encircle zero or two CIs and do not.

We consider a two-level atomic or molecular system with ground and
excited states $|g\rangle$ and $|e\rangle$, with zero-field energy
separation $\hbar\omega_0$, coupled by a standing-wave optical field
with frequency $\omega_{\rm L}/2\pi$ and a node at $y=0$. An
inhomogeneous magnetic field $B_z(x)$ is oriented along the laboratory
$z$ axis and varies along the $x$ axis. The dressed-state Hamiltonian
is
\begin{equation}
\left(\begin{array}{cc}
\hbar\omega_{\rm L} +
\mu_gB_z(x) & (\hbar\Omega_{eg}/2)\sin ky \\
(\hbar\Omega_{eg}/2)\sin ky &
\hbar\omega_0
+\mu_eB_z(x) \end{array} \right),
\label{eq:ham}
\end{equation}
where $\Omega_{eg}$ is the Rabi frequency of the transition and
$k=\omega_{\rm L}/c$. If the optical field is resonant with the
transition between the two states at a magnetic field
$B_z(x=0)=B_0$, the eigenvalues of (\ref{eq:ham}) will form a
seam of CIs along a line at (0,0,$z$). If the atomic or
molecular cloud is large enough, there will be additional seams
of CIs at regular intervals such that $y_n=n\pi c/\omega_{\rm
L}$ for integer $n$.

CIs of this type can in principle occur for any optical
frequency. However, the sharpness of the excited state is
limited by spontaneous emission, which is likely to limit the
use of visible and even near-infrared transitions because of
the $\omega_0^3$ factor in the equation that relates the
Einstein $A$ coefficient to the transition dipole. This might
be overcome for strongly forbidden transitions, but then the
Rabi frequencies would be quite low. We will return to the
possible use of infrared transitions below, but we will begin
by considering CIs based on microwave transitions.

An alkali-metal atom with nuclear spin $I$ has two zero-field hyperfine
states with $F=I\pm\frac{1}{2}$, each of which splits into $2F+1$
sublevels in a magnetic field. The two states with the same $M$ but
different values of $F$ have equal and opposite Zeeman effects. There
are magnetic-dipole-allowed microwave transitions between states
($F,M$) and ($F$+1,$M'$) with $\Delta M$=$M'$--$M$=0,$\pm1$. As an
example, we consider $^{87}$Rb, which has $I=\frac{3}{2}$ and a
zero-field splitting of 6.835 GHz between its $F=1$ and $F=2$ states.
Let us consider a central static field $B_0=200$~G at $y=0$, so that
the splitting between the ($F,M$) = (1,+1) and (2,+1) sublevels is
7.115~GHz. The transition between these states has $\Delta M=0$ so
requires an oscillating magnetic field along the $z$ axis. To provide
this, consider placing the atoms in a rectangular or cylindrical
microwave cavity in which the transverse electric mode TE$_{011}$ is
excited. This creates magnetic field lines that circle in the $yz$
plane, with a node in $B_z(\omega_Lt)$ at $y=0$.

To create an observable CI in a BEC, it seems reasonable to aim
for a ring radius of 6~$\mu$m. This could be achieved with a
magnetic field gradient $dB_z/dx$ of 50 G/cm and an optical
trap with a force constant of 28 nK/$\mu$m$^2$. To balance the
Zeeman effect would require an optical Zeeman effect of the
same magnitude, i.e. about 1 $\mu$K at $y=6$~$\mu$m,
corresponding to $\Omega_{eg}\sin ky=21$~kHz at $y=6$~$\mu$m or
33~MHz at the magnetic field antinode for a cubic cavity. This
is a realistically achievable field, and indeed Spreeuw {\em et
al.}\ \cite{Spreeuw:1994} in 1994 demonstrated trapping of
ground-state Cs atoms in a microwave trap with a central Rabi
frequency of 36~MHz in a spherical cavity, despite a relatively
low cavity Q-factor of 5500 due to the need for holes to allow
access for atomic and laser beams.

\begin{figure}
\begin{center}
\includegraphics[width=0.28\textwidth]{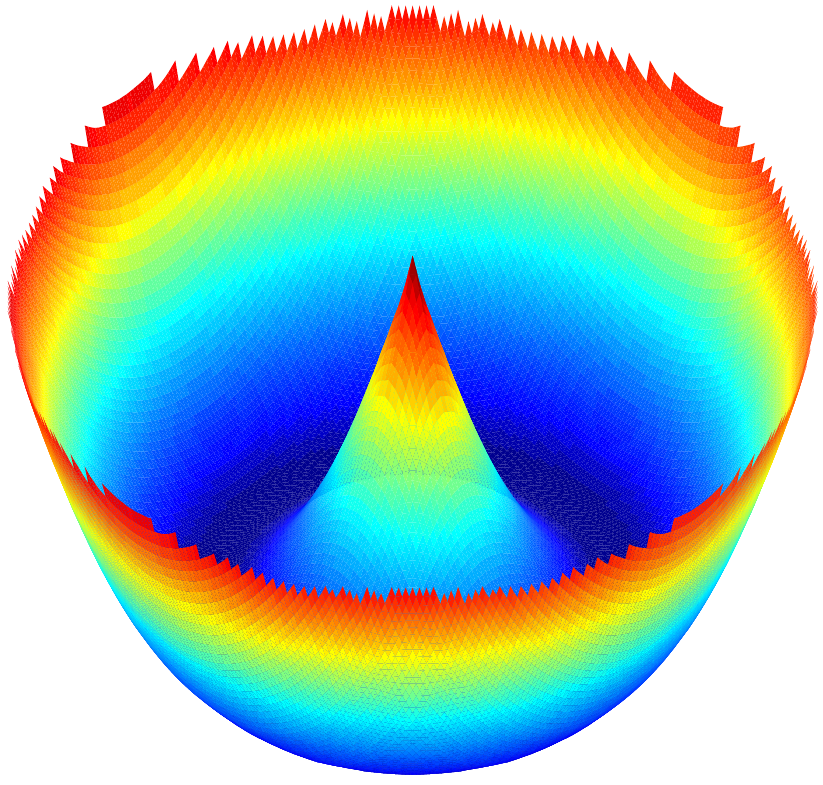}
\includegraphics[width=0.50\textwidth]{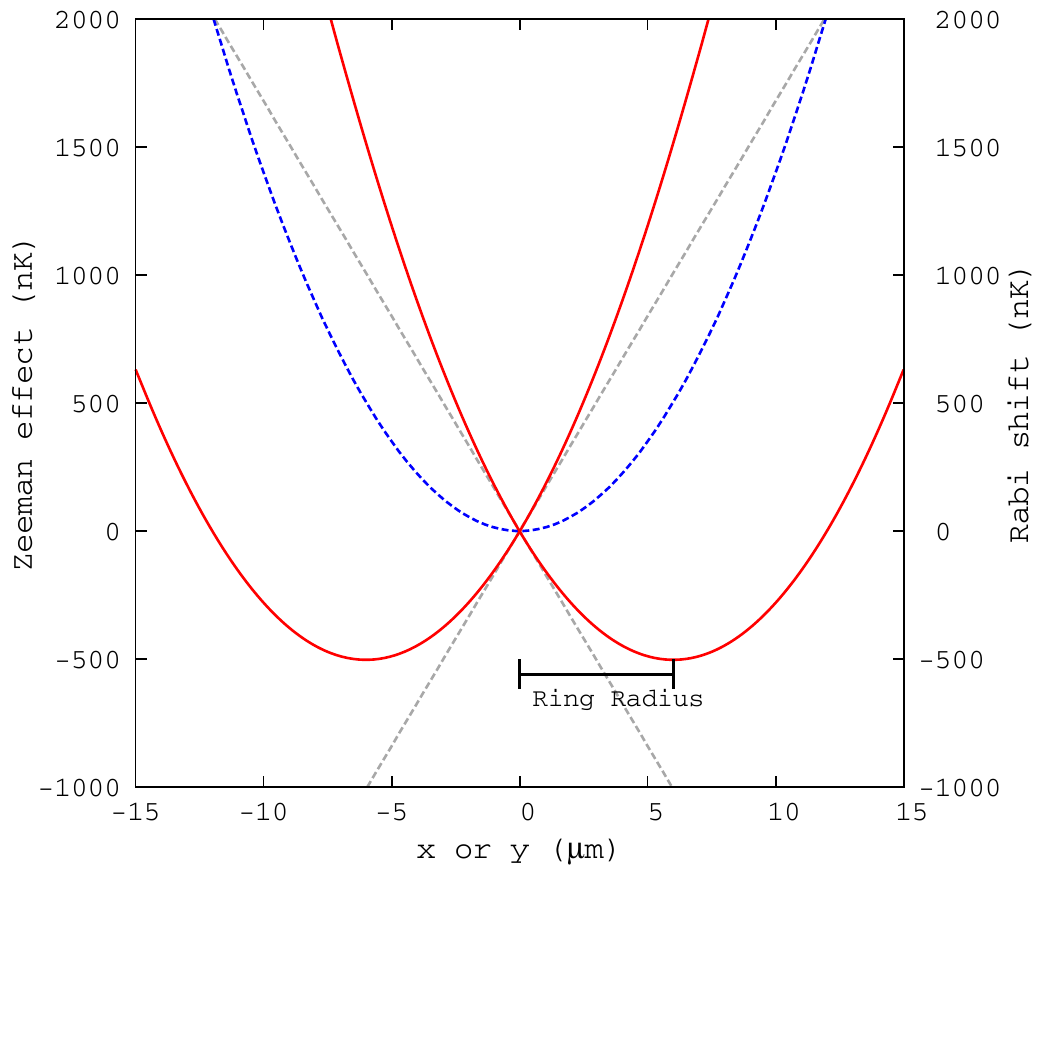}
\vspace{-2.0cm}
\caption{(color online). Upper panel: toroidal potential formed
by creating an optically induced conical intersection at the
centre of an optical dipole trap; lower panel: contributions of
the individual contributions to the toroidal potential for
$^{87}$Rb for the parameters given in the text.}
\label{fig:CItrapped}
\end{center}
\end{figure}

Such a trap would create a potential minimum around a ring in
the $xy$ plane, about 0.5 $\mu$K below the energy of the CI, as
shown in Fig.\ \ref{fig:CItrapped}. Rubidium atoms are far too
massive to tunnel into such a potential and reach the central
point. There is no point at which the magnetic field is zero,
so losses due to Majorana spin flips should not be a problem.
The arrangement described thus far is cylindrically
symmetrical, and confinement in the $z$ direction must be
provided optically, since the trapped atoms are in a
combination of two states with different Zeeman effects and
cannot be levitated magnetically.

The quantum behaviour of an atom or a BEC in a trap of this type is
exactly the same as discussed by Wallis {\em et al.}\
\cite{Wallis:PRL:CI:2009} for a CI induced by static fields. For a
single particle, the Berry phase due to the CI produces antiperiodic
boundary conditions $\Phi(\phi+2\pi)=-\Phi(\phi)$ for the motion around
the intersection, where $\phi$ is the angle in the $xy$ plane. If the
ring is sufficiently flat, the Berry phase produces states with
half-integer rotational angular momentum around the intersection. This
differs from the Berry phases that can exist in quadrupole magnetic
traps or storage rings \cite{Ho:vortex:1996, Olshanii:1998,
Zhang:ring:2006}, because in our case the magnetic field does not vary
in direction around the ring. For a BEC that can be modeled by the
Gross-Pitaevskii equation, the self-energy is itself anisotropic
because the molecular eigenstate changes from $|g\rangle$ to
$|e\rangle$ around the ring. The ground state is a flowing state with
half-integer angular momentum if the average self-energy is large
enough to overcome its own anisotropy and that of the potential
\cite{Wallis:PRL:CI:2009}. $^{87}$Rb is a particularly favorable case
in this respect, because its singlet and triplet scattering lengths are
both positive and quite similar to one another, so that all its
hyperfine states have quite similar scattering lengths.

Several variants of this arrangement can be envisaged. CIs
could be created for a variety of different atoms and pairs of
atomic states. For microwave transitions of alkali metal atoms,
however, transitions with $\Delta M=0$ are optimum because the
two states have equal and opposite Zeeman effects, so form an
untilted CI. A small difference in Zeeman effects can be
compensated by an offset of the optical potential, as in the
static case \cite{Wallis:PRL:CI:2009}, but it will be much
harder to create a flat ring if the two crossing states have
substantially different Zeeman effects.

It might be possible to generate microwave fields of sufficient
power on an atom chip, using microwave near-fields, such as
have been used for the coherent manipulation of ultracold atoms
on atom chips \cite{Treutlein:2008, Bohi:2009}. Two parallel
current-carrying wires along $x$, with in-phase currents at
microwave frequencies, will create an oscillating magnetic
field with a nodal plane in the magnetic field perpendicular to
the chip $B_z(y)$ half-way between the wires, in the $xz$ plane
at $y=0$. In this case confinement might be achieved with an
optical trap created by reflecting a laser beam from the chip
surface \cite{Gallego:2009}.

It is also conceivable that a CI might be created using
radiofrequency fields to couple two atomic states of the same
$F$ but different $M$. In this case the requirement for nearly
equal and opposite Zeeman effects is met at low fields by
states with equal and opposite $M$ values, and together with
the selection rule $\Delta M=\pm1$ this requires
$M=\pm\frac{1}{2}$ for a 1-photon transition. Such states exist
only for fermionic atoms. However, another possibility would be
to use high-field states, with magnetic moments determined
mostly by the electron spin $M_S$ rather than the total spin
$M$. Strong radiofrequency fields have been generated on atom
chips and used to shape trap potentials \cite{Zobay:2001,
Colombe:2004}, and even to generate ring-shaped traps
\cite{Heathcote:2008}. However, even if a node in a
radiofrequency field can be positioned accurately enough, it
will be challenging to produce the oscillating fields needed to
form a CI within a few $\mu$m of the nodal plane.

There is a final possibility that warrants consideration, though it is
further from implementation than the atomic examples considered above.
If light of much shorter wavelength is used to provide the optical
coupling, it is possible to envisage an {\em array} of CIs within a
single BEC, which would allow the study of even richer interference
effects. To achieve this, the spacing between intersections must be
smaller than the size of the trapped ultracold gas. This requires a
laser half-wavelength $\lambda_{\rm L}/2=\pi c/\omega_{\rm L}$ less
than a few hundred $\mu$m. Furthermore, it would be necessary to create
an array of individual optical traps to surround each CI separately,
with a controllable barrier between them. This probably requires the
intersections to be separated by at least a few $\mu$m. Such traps
cannot be created with the near-resonant laser that generates the CIs,
but it might be achieved using a holographically generated array of
microtraps \cite{Bergamini:2004} centered on the intersections.

\begin{figure}
\begin{center}
\includegraphics[width=0.4\textwidth]{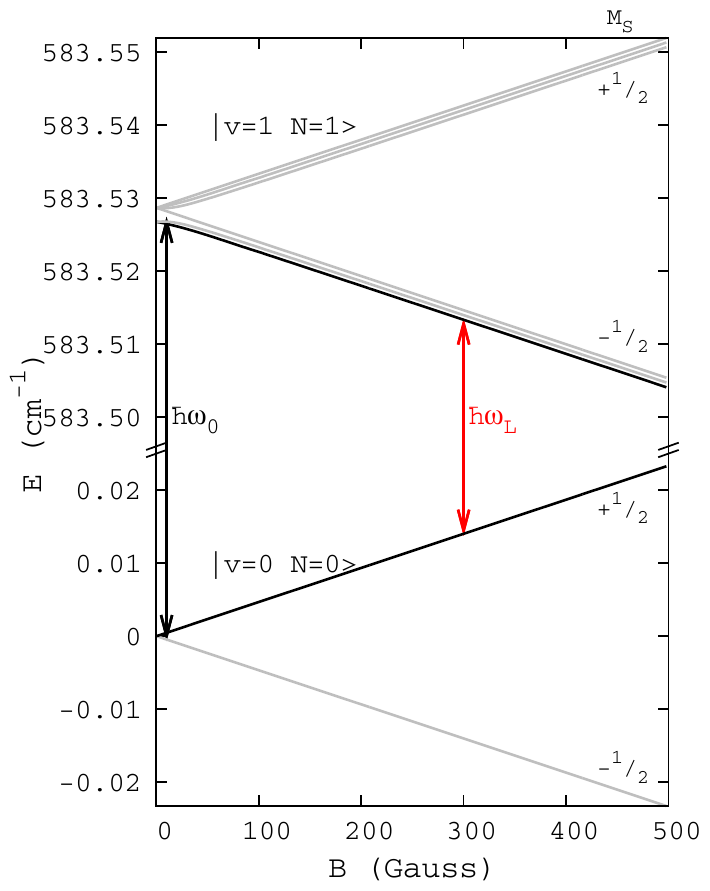}
\caption{(color online). Magnetic field dependence of the
$|v=0,N=1\rangle$ and $|v=1,N=1\rangle$ states of
CaF(X$^2\Sigma^+$). To form an array of CIs
the $|v=0;N=0,M_N=0,M_S=\frac{1}{2}\rangle$ and
$|v=1;N=1,M_N=1,M_S=-\frac{1}{2}\rangle$ states (black), with a
zero field separation of $\hbar\omega_0$, are coupled by a
linearly polarized laser standing wave $\hbar\omega_{\rm L}$
that is resonant with the two-states at $B_0=300$
G.}\label{fig:CaFstates}
\end{center}
\end{figure}

An array of CIs might in principle be created for either atoms
or molecules, but we are not aware of a coolable atom with a
suitable transition in the required frequency range. However,
the wavelength range of 3 to 300 $\mu$m corresponds to a level
separation of 30 to 3000 cm$^{-1}$, which is typical of
molecular vibrations. We therefore consider the example of
CaF(X$^2\Sigma^+$), which has a vibrational frequency of 588
cm$^{-1}$ (17 $\mu$m). CaF is not yet available ultracold,
though it has been produced at mK temperatures by buffer-gas
cooling \cite{Maussang:2005}. The levels involved in the $(v,N)
= (0,0) \leftrightarrow (1,1)$ transition of CaF are shown in
Figure~\ref{fig:CaFstates} as a function of magnetic field. At
high magnetic field all the rovibrational states are
characterized by the spin-projection quantum number
$M_S=\pm\frac{1}{2}$.

We consider a CI formed from the two states that are
predominantly $|v=0;N=0,M_S=+\frac{1}{2}\rangle$ and
$|v=1;N=1,M_N=1,M_S=-\frac{1}{2}\rangle$, coupled by a laser
linearly polarized along the magnetic field axis $z$. These two
states are shown in black in Figure~\ref{fig:CaFstates}. The
transition dipole moment between the $v=0$ and 1 states can be
estimated from the gradient of the permanent electric dipole
moment of the molecule at its equilibrium bond length
\cite{Langhoff:1986}. Assuming that the vibrational
wavefunctions are harmonic, the transition dipole moment can be
estimated as 0.25~D. The natural lifetime of the excited state
is $\tau =1/A_{eg}\approx 0.25$~s, so that the width due to
spontaneous emission is negligible.

\begin{figure}
\begin{center}
\includegraphics[width=0.45\textwidth]{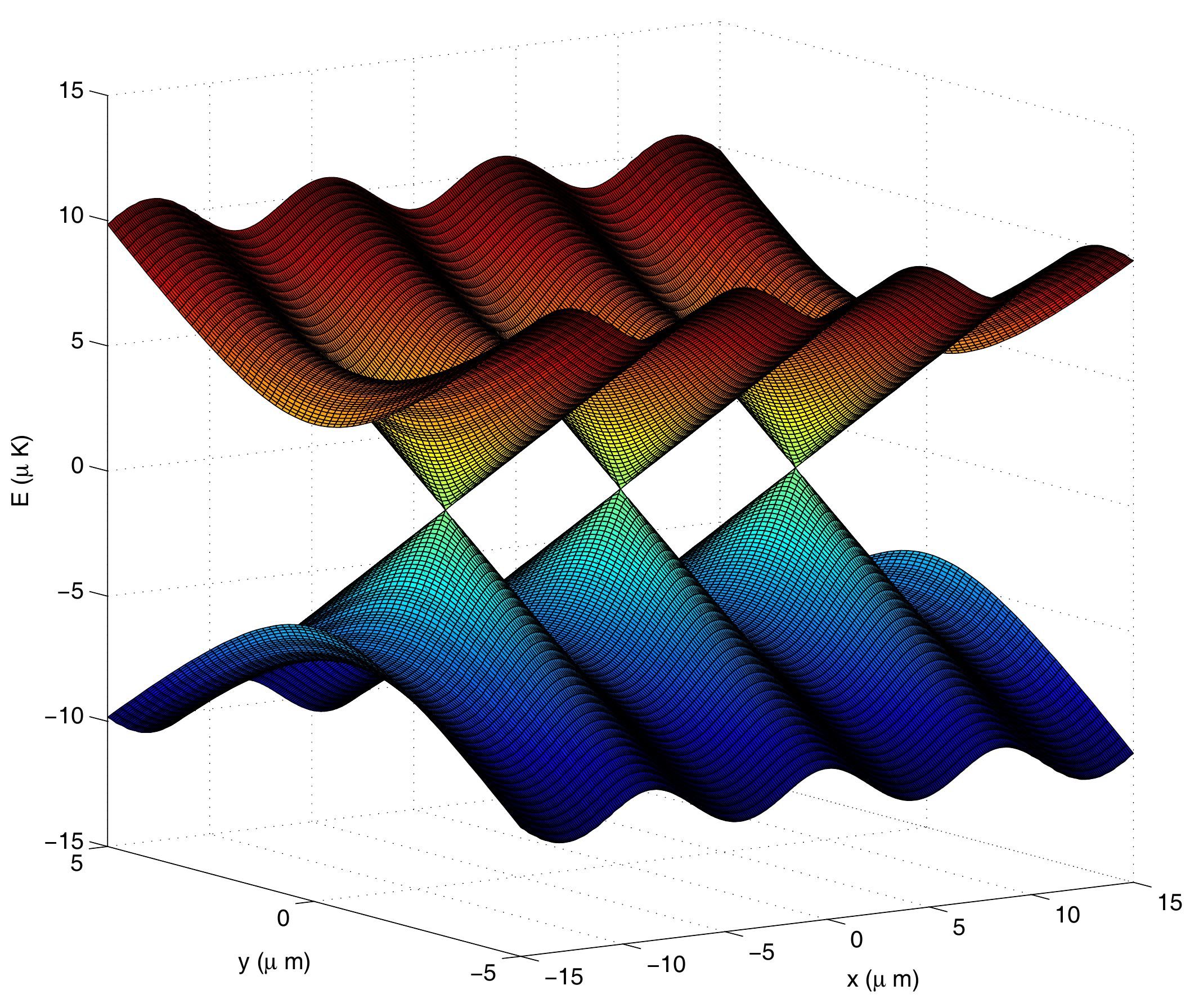}
\caption{(color online). Array of conical intersections formed
by a 5~mW/cm$^{2}$ laser standing wave resonant with two
ro-vibrational states of CaF, with a central magnetic field
$B_0$=300~G and a constant magnetic field gradient
$dB_z/dy=553$~G/cm.} \label{fig:CIarray}
\end{center}
\end{figure}

The dressed-state Hamiltonian is constructed with a field-free energy
separation $\hbar\omega_0 = 583.5\ hc$ cm$^{-1}$, and a standing-wave
laser with an intensity of 5 mW/cm$^2$, resonant at a magnetic field
$B_0=300$~G. The intersecting dressed-state eigenvalues are shown in
Figure~\ref{fig:CIarray}. CIs occur every $\lambda_{\rm L}/2\approx  9
~\mu$m, at each node in the standing wave.

In conclusion, we have demonstrated that conical intersections
may be created as a function of laboratory position space by
combining an optical field and a static inhomogeneous magnetic
field. This may be achieved for pairs of states of either the
same or different parity, which are shifted into near-resonance
by the optical field and then tuned across one another with the
magnetic field. Such a CI can be created for atoms such as
$^{87}$Rb using realistically achievable microwave fields. The
resulting Berry phase produces antiperiodic boundary conditions
for states that encircle the CI, and may result in flowing
states with half-integer angular momentum. At higher
frequencies, optical fields might be used to produce arrays of
CIs, each confined in a microtrap and interacting with one
another.

The authors are grateful to EPSRC for a research studentship
for AOGW and for funding the collaborative project QuDipMol
under the ESF EUROCORES program EuroQUAM. We also acknowledge
valuable discussions with Charles Adams, Simon Gardiner, Sam
Meek, Liam Duffy and Gerard Meijer.

\bibliography{../../all,thref}

\end{document}